Scaling the energy conversion rate from magnetic field reconnection to different bodies
F. S. Mozer and A. Hull
Space Sciences Laboratory, University of California, Berkeley, CA. 94720

ABSTRACT

Magnetic field reconnection is often invoked to explain electromagnetic energy conversion in planetary magnetospheres, stellar coronae, and other astrophysical objects. Because of the huge dynamic range of magnetic fields in these bodies, it is important to understand energy conversion as a function of magnetic field strength and related parameters. It is conjectured theoretically and shown experimentally that the energy conversion rate per unit area in reconnection scales as the cube of an appropriately weighted magnetic field strength divided by the square root of an appropriately weighted density. With this functional dependence, the energy release in flares on the Sun, the large and rapid variation of the magnetic flux in the tail of Mercury, and the apparent absence of reconnection on Jupiter and Saturn, may be understood. Electric fields at the perihelion of the Solar Probe Plus mission may be tens of volts/meter.

I INTRODUCTION

Magnetic field reconnection occurs when two magnetized plasmas, having a sheared magnetic field across their interface, flow towards each other[1]. Its characteristic feature is a modification of the original magnetic field topology that results from the presence in some relatively small region of dissipative processes that convert electromagnetic energy to plasma energy[2,3]. Plasma from one of the inflowing regions becomes magnetically connected to plasma from the other inflowing region as a result of the topological change due to reconnection. The magnetic field geometry of reconnection is illustrated in Fig. 1, in which a thin sheet of current (the grey rectangle of Fig. 1), flowing into the plane of the figure (in the +Y-direction), is associated with the shear in the magnetic field component, $B_Z$. The X-direction is normal to the thin current sheet and the Z-direction is that of the reconnecting magnetic field components that combine to form the reconnected magnetic field lines above and below the X-line, at the center of the figure.

Magnetic field reconnection may be invoked to explain cosmic rays with energies as great as $10^{20}$ eV[4] as well as the 18-order-of-magnitude range of energy release between laboratory machines, the terrestrial magnetosphere, the Sun, Galactic diffuse X-ray emission, Radio Lobes and the Crab Nebula[4,5]. To understand this tremendous dynamic range, it is necessary to know the dependence of a single reconnection event on the magnetic field strength and associated parameters. In Sec. II, it is conjectured that the energy conversion rate per unit area in a single reconnection event scales as the cube of an appropriately weighted average magnetic field strength divided by the square root of an appropriately weighted density. In Sec. III, it is conjectured that $E_X = B_1(B_1+B_2)/80.38 dn_0^{0.5}$ where $B_1$ and $B_2$ are the magnetic fields on the two sides of the current sheet, $n_0$ is a weighted plasma density and d is the current sheet thickness in units of the ion skin depth. In Sec. IV, Polar Satellite observations of sub-solar reconnection, occurring over a factor of 10 in magnetic field strengths, are analyzed to show that this functional form for $E_X$ provides a better fit to the experimental data than do modified functions of B and $n_0$, and that $R^2$, the fraction of the total squared error explained by this functional form, is a surprisingly large 0.90. Sec. V utilizes the scaling relationships to discuss reconnection in the solar system and to show that, on the Sun, a single reconnection event, occurring over a reasonable volume, may release enough energy to explain solar flares; at Mercury it may account for flux pileup in the tail; and at Jupiter and Saturn reconnection is probably negligible. Sec. VI briefly comments on scaling in the astrophysical context..

II CONJECTURE THAT $E_Y$ SCALES AS $B_0^2/n_0^{0.5}$ AND THE ENERGY CONVERSION RATE SCALES AS $B_0^3/n_0^{0.5}$

In the following discussion, it is initially assumed that the reconnection geometry is two-dimensional, that the plasma densities and the magnitudes of $B_Z$ at the right and left sides of Figure 1 are the same, and that $B_Y = 0$. In the presence of an into-the-plane electric field ($E_Y > 0$), the plasma flows toward the center of the figure from each side at the velocity $\mathbf{E}\mathbf{X}\mathbf{B}/B^2 = E_Y/B_Z$. This speed may be normalized by the Alfven speed, $V_A = [B^2/\mu_o mn]^{0.5}$, where m is the ion mass, to form the dimensionless reconnection rate, defined as

Reconnection Rate $\equiv E_Y/(B_Z V_A)$ (1)

In many simulations[8] and measurements[9,10], the reconnection rate is constant within a factor of two or so at a value ~0.1. It is important to note that this is a measure of the speed of the inflowing plasma and magnetic field but it is **not** a measure of the number of field lines that reconnect per second or the energy conversion rate.

In accordance with simulations and measurements, we assume that the reconnection rate is constant at the value, M = 0.1, in all reconnecting plasmas. In this case, equation (1) becomes

$E_Y = M V_A B_Z \; \alpha \; B_Z^2/n^{0.5}$ (2)

This dependence has also been found in computer simulations[11] but has never before been tested against experimental data.

The magnetic flux that reconnects per second per unit length is the field line inflow speed, $\mathbf{E}\mathbf{X}\mathbf{B}/B^2$, times the density of field lines, B, which, for the present case, simplifies to being $E_Y$. Thus, the flux reconnection rate is proportional to $B_Z^2/n^{0.5}$, which is more than $10^7$ times greater in the solar corona than at the Earth's sub-solar magnetopause, in spite of the fact that the reconnection rate, defined by equation (1), is the same at the two locations if M = 0.1.

To make quantitative comparisons of the measured and expected electric fields in asymmetric reconnection, it is necessary to modify equation (2) to accommodate the more general case of different plasma densities and reconnecting magnetic field magnitudes at the two sides of the reconnection region. In this case[6,7], $B_Z$, n and $V_A$ are replaced by

$$B_0 = 2B_1B_2/(B_1+B_2)$$

$$n_0 = (n_1B_2+n_2B_1)/(B_1+B_2)$$

and

$$V_0 = [B_1B_2/\mu_o m n_0]^{0.5}$$

where $B_1$ and $B_2$ are absolute values of the reconnecting magnetic field strengths on the two sides of the current sheet, as illustrated in Fig. 1, and $n_1$ and $n_2$ are the plasma densities on the two sides. Thus, equation (2) for asymmetric reconnection becomes

$$E_Y = MV_0B_0 \; \alpha \; B_0(B_1B_2)^{0.5}/n_0^{0.5} \qquad (3)$$

This is the quantitative expression for $E_Y$ that will be used for making comparisons with experimental data in the terrestrial magnetosphere.

Because the energy conversion rate per unit area is proportional to the Poynting vector, **E**X**B**/$\mu_o$, entering on each side of the current sheet,

$$\text{Energy conversion rate per unit area} \; \alpha \; MV_0B_0(B_1+B_2)/\mu_o \qquad (4a)$$

For scaling to other objects where the reconnection magnetic fields and plasma densities on the two sides of the current sheet are not known, the dependences on these parameters will be ignored and the general scaling law that will be invoked is

$$\text{Energy conversion rate per unit area} \; \alpha \; B^3/n^{0.5} \qquad (4b)$$

III CONJECTURE THAT $E_X$ IS PROPORTIONAL TO $B_1(B_1+B_2)/n_0^{0.5}$

The Generalized Ohm's law is

$$\mathbf{E}+\mathbf{U}_I\mathbf{x}\mathbf{B}=\mathbf{j}\mathbf{x}\mathbf{B}/en-\nabla\cdot\mathbf{P}_e/en-(m_e/e)[\partial\mathbf{U}_e/\partial t + (\mathbf{U}_e\cdot\nabla)\mathbf{U}_e]+ \eta\mathbf{j} \qquad (5)$$

where $\mathbf{U}_I$ is the ion flow speed, $\nabla\cdot\mathbf{P}_e$ is the divergence of the electron pressure tensor and $\eta$ is the resistivity. In the ion diffusion region of a collisionless reconnection event, the ion flow is small and the second, third, fourth and fifth terms on the right are negligible[12]. In this case

$$E_X = j_Y B_Z/en \tag{6}$$

The out-of-plane current, $j_Y$, is given by Ampere's law as $(B_1 + B_2)/\mu_o\delta$ where $\delta$ is the thickness of the current sheet in the X-direction. Expressing $\delta$ in units of the ion skin depth, $c/\omega_{pI} = c(ne^2/\varepsilon_0 m)^{-0.5}$ gives

$$d = \delta/(c/\omega_{pI}) \tag{7}$$

where d is the current sheet thickness in units of $c/\omega_{pI}$. Many simulations have shown that d is independent of the magnetic field (as assumed in the present analysis) and is the order of one[3]. Allowing $B_Z$ in equation (6) to become $0.5B_1$ because the large $E_X$ value is found on the magnetospheric side of the magnetopause at a location where the magnetic field is about half of the peak value[12], and allowing the density to be $n_0$ because the density observed in regions where equation (6) is obeyed is intermediate between the densities on the two sides of the current sheet[12], the expression for $E_X$ in mV/m, B in nT, and $n_0$ in $cm^{-3}$ becomes

$$E_X = B_1(B_1 + B_2)/(80.38\ dn_0^{0.5}) \tag{8}$$

This is the quantitative expression for $E_X$ that will be used for comparison with experimental data in the terrestrial magnetosphere.

IV EXPERIMENTAL DATA

Equations (3) for $E_Y$ and (8) for $E_X$ have been tested against Polar satellite data, an example of which, in Figure 2, shows a sub-solar magnetopause crossing at a geocentric distance of 5.3 Earth radii, which is the closest magnetopause event to the Earth recorded in the 13 year Polar mission. Because the dipole magnetic field of the Earth varies inversely as the cube of the radial distance, the expected magnetospheric magnetic field should be about $(9/5.3)^3 \sim 5$ times larger than the typical 50 nT field observed for reconnection at the typical 9 Earth radius event distance observed on Polar. As shown in panel (a), the observed reconnection field agrees with this extrapolation, having an average value of 230 nT in the magnetosphere, to the left of Figs.1 and 2. The average field in the magnetosheath during the final 30 seconds of Fig. 2 is -161 nT.

For plasma densities of 7 and 16 $cm^{-3}$ in the magnetosphere and the magnetosheath, for a current sheet thickness of two ion skin depths (as discussed below), and for $B_1 = 250$ nT and $B_2 = -161$ nT, equation (8) gives $E_X = 189$ mV/m whereas the measured average value is 162 mV/m. For M = 0.1, equation (3) predicts $E_Y = 17$ mV/m, which compares with the four-second-averaged $E_Y$ of 24 mV/m. The uncertainty in the average $E_Y$ is the same order as the measured value because:

1. Temporal fluctuations of $E_Y$ during different time intervals over which it may be averaged can change the result by a factor of two.

2. The uncertainty in $E_Y$ associated with rotation of the data into the minimum variance frame is similar to the magnitude of $E_Y$ because $E_X$ is an order-of-magnitude larger than $E_Y$.
3. The translation of the data along the X-axis into the frame of the magnetopause changes $E_Y$ by an amount comparable to its value, which has not been done for the data of Figure 2 because the magnetopause speed over the spacecraft is not known.

For these reasons $E_Y$ cannot be used to test equation (3) and we shall use the measured $E_X$ to test equation (8). The uncertainty of $E_X$ does not depend on problems 2 and 3 above, but selection of the averaging interval does affect the result. The -161 nT average of the last thirty seconds of data in Fig. 2a was used as the magnetosheath $B_1$ in the following analysis because this was its average value after the illustrated time interval. However, the uncertainty in this parameter can be typically ~25%, which must be remembered when evaluating the following statistical analyses.

Forty eight reconnection events were collected during 2001-2003 and at other times of major magnetic storms when the magnetic local time at the spacecraft was between 0900 and 1500 and the magnetic latitude was less than 35°. $E_X$ for each of these events is plotted versus equation (8a) with d = 1 in panel (b) of Fig. 3. The 16 events having plasma outflow velocities >300 km/sec are described in panel (a) of this figure. These 16 events are considered to be more reliable than the remaining 32 events because the spacecraft may have been closer to the diffusion region and active reconnection during each of them. The largest electric field event in Fig. 3 is the event shown in Fig. 2.

For Fig. 3a, the correlation coefficient, $R^2$, which is the fraction of the total squared error explained by equation (8a), is a surprisingly large 0.90. This and the $R^2$-value of 0.84 in panel (b) strongly support the conclusion that the magnetic flux reconnection rate and the energy conversion rate depend on powers of the reconnection magnetic field.

The functional dependence of the electric field on the magnetic field and the plasma density has been studied by computing the correlation coefficients of the least squares fits to different functions of B and $n_0$, as given in Table 1. Examination of these $R^2$-values shows that:
- the best overall fit to $E_X$ is given by the functional form $B_1(B_1+B_2)/n_0^{0.5}$
- the $B_1(B_1+B_2)$ dependence provides a better fit than does either a $[B_1(B_1+B_2)]^{0.5}$ or $[B_1(B_1+B_2)]^{1.5}$ dependence,
- including the plasma density provides better fits than not including it.

It is noted that the value of $R^2$ for 16 events and a functional dependence proportional to $[B_1(B_1+B_2)]^{0.5}$ is large relative to the other values. This is because B and $n_0$ roughly scale together, so a B dependence is not greatly different from a $B^2/n_0^{0.5}$ dependence.

The current sheet thickness found from the slope of the least squares fit of Fig. 3a is about two ion skin depths. This result is consistent with other estimates[3] and adds validity to the entire procedure.

The dependences of the measured and scaled electric fields were compared with the magnitude of the measured guide magnetic field (which ranged from 0.1 to 1.1) and no relationship was found.

V APPLICATION TO THE SOLAR SYSTEM

During its impulsive phase, a solar flare may release as much as $10^{25}$ Joules of energy in ~1000 seconds ($10^{22}$ W). This phase is often followed by a gradual phase involving helmet streamers that have a reconnection-like magnetic field geometry, from which ~$10^{23}$ Joules may be released. Surprisingly, as much as 50% of this energy ends in accelerated electrons[13]. It is often assumed that these $10^{20}$-$10^{22}$ Watts are released by reconnection in the solar corona. Assuming a magnetic field of 100 Gauss and a plasma density of $10^8$ cm$^{-3}$ (each of which could be an order-of-magnitude different) at a reconnection site in the solar corona[14], the energy release per unit area from equation (4a) is $10^7$ W/m$^2$, which is a factor of $10^{11}$ greater than that at the terrestrial magnetopause. If the power of $10^7$ W/m$^2$ is converted over an area of 20,000 by 20,000 km$^2$ (which is $10^6$ proton skin depths by $10^6$ proton skin depths or 0.01% of the solar surface area), the electromagnetic energy released by one reconnection event is sufficient to power solar flares and helmet streamers. Addition of multiple reconnection sites and the formation of islands that coalesce to release further energy can reduce the required area by a large factor[15,16]. The benefit of the scaling law is that the requirement on the area of the reconnection site or the multiple reconnection and island formation is diminished by a factor of ~$10^{10}$ from that required in the absence of the scaling.

At the 9.5 solar radius perihelion of the Solar Probe Plus satellite, the convection electric field can be tens of volts/meter and the convective flow can be thousands of km/sec. Instruments must be designed with these dynamic ranges in mind.

Messenger satellite observations of Mercury's magnetotail[17] have shown that marked increases of the tail field are probably caused by enhanced reconnection and that the magnetic energy content of the tail increased at a rate greater than 5 times that at similar events in the terrestrial magnetotail. Equation (3) states that the flux loading per unit time is proportional to $B_0^2/n_0^{0.5}$. Although the plasma density at Mercury was not measured, the fact that the observed magnetotail field at Mercury was ≥5 times larger than that in the terrestrial magnetotail suggests that the Mercury tail observations were associated with enhanced flux reconnection on the dayside due to the larger solar magnetic field in its vicinity.

The role of magnetic field reconnection at Jupiter has been a topic of much discussion, with views ranging from reconnection producing an open magnetosphere, to a viscous interaction along the magnetopause which, in combination with the large internal magnetic field, internal plasma sources, and fast rotation, dictates the closed magnetospheric dynamics[18]. The uncertainties of the magnetic fields and densities at the sub-solar Jovian magnetopause allow only a qualitative estimate of the electromagnetic energy conversion rate per unit area due to reconnection. For a reconnecting magnetic field of 1-2 nT and a plasma density of ~1 cm$^{-3}$, the energy conversion rate is more than 10,000 times smaller at Jupiter than at the terrestrial magnetopause. This supports the view that reconnection may be unimportant compared to other processes responsible for auroral activity on Jupiter, as has been noted [19].

It has been shown that the intensity of Saturn's auroral emissions depends on solar wind pressure and internal processes[19,20], so reconnection may not be important. This may be explained by the fact that the $B^3/n^{0.5}$ energy conversion rate factor is at least 15,000 times smaller for Saturn than for the Earth.

VI APPLICATION TO HIGH ENERGY ASTROPHYSICS

. Because astrophysical objects are observed to have maximum particle energies that depend on a power of the magnetic field[5] and because the acceleration mechanisms that produce them are not well understood[4,21], it would be desirable to extrapolate the scaling laws of this paper to such objects. If the electric field scaling in such regimes is also proportional to the Alfven speed (equal to the speed of light) times B, which is a scaling of energy conversion per unit area proportional to $B^2$, then a significant energy release from reconnection could be achieved in astrophysical objects.

**Acknowledgements**

This work was supported by NASA Grant NNX09AE41G-1/11. The authors thank Professor Fran Bagenal, Dr. Hugh Hudson and Professor Michael Shay for helpful comments and information.

**TABLE 1   $R^2$ for different functions of B and $n_0$**

|  | 16 Events | 48 Events |
|---|---|---|
| $[B_1(B_1+B_2)]^{0.5}$ | 0.83 | 0.59 |
| $[B_1(B_1+B_2)]^{1.0}$ | 0.74 | 0.68 |
| $[B_1(B_1+B_2)]^{1.5}$ | 0.40 | 0.36 |
| $[B_1(B_1+B_2)]^{0.5}/n_o^{0.5}$ | 0.49 | 0.37 |
| $[B_1(B_1+B_2)]^{1.0}/n_o^{0.5}$ | 0.90 | 0.84 |
| $[B_1(B_1+B_2)]^{1.5}/n_o^{0.5}$ | 0.58 | 0.56 |

FIGURE CAPTIONS

Figure 1. The magnetic field geometry for reconnection resulting from two plasmas flowing towards each other with a sheared magnetic field at their interface.

Figure 2. Electric fields in millivolts/meter and magnetic fields in nanoTesla obtained on the Polar Satellite during a four minute interval in which the magnetopause was crossed at 1040 magnetic local time and 14° magnetic latitude. The data has been rotated into the minimum variance coordinate system. Note that the amplitude scales for $E_Y$ and $E_Z$ in panels (c) and (d) differ by a factor of two from that for $E_X$ in panel (b). The magnetosphere is at the left of the figure with its positive $B_Z$ in panel (a) and the magnetosheath, on the right, has a negative reconnection magnetic field.

Figure 3. Linear least squares fits of $E_X$ to equation (8) with d = 1, measured at 48 sub-solar magnetopause crossings in panel (b) and to the 16 of these crossings having a plasma outflow speed greater than 300 km/sec in panel (a).

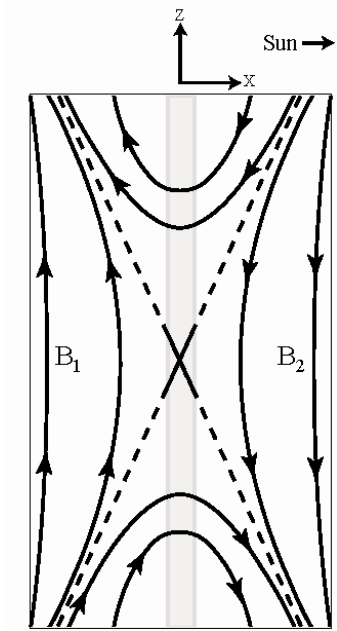

Figure 1

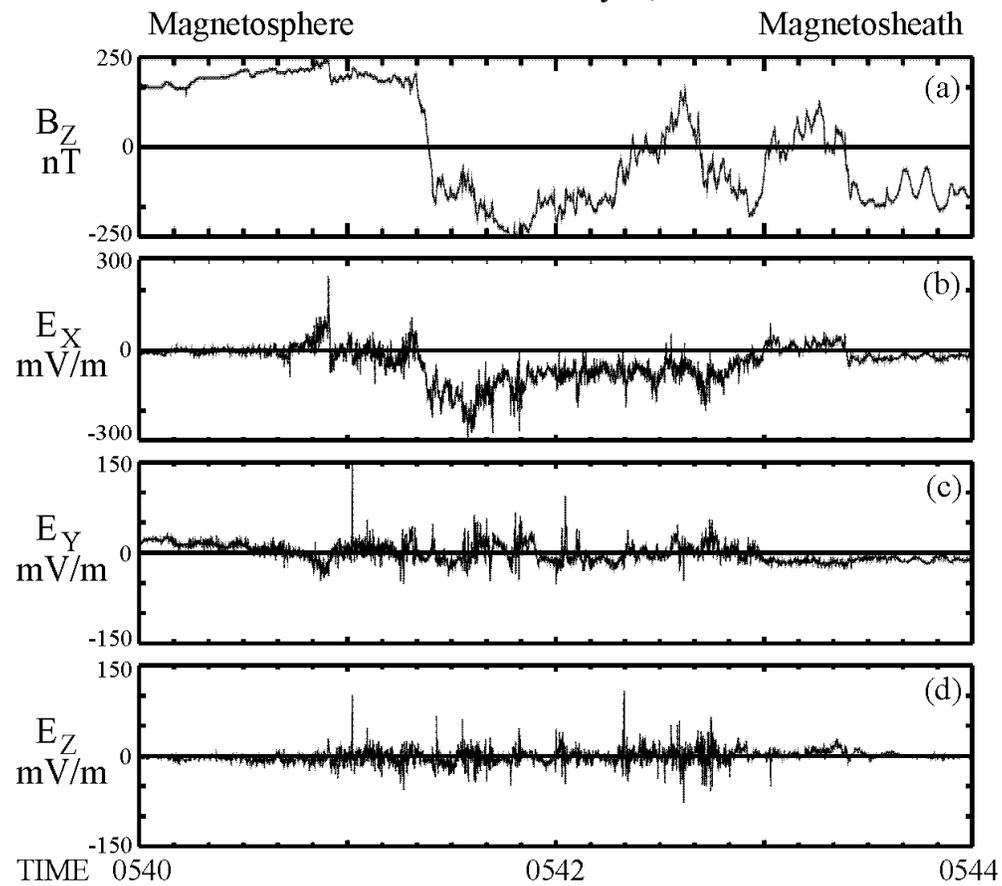

Figure 2

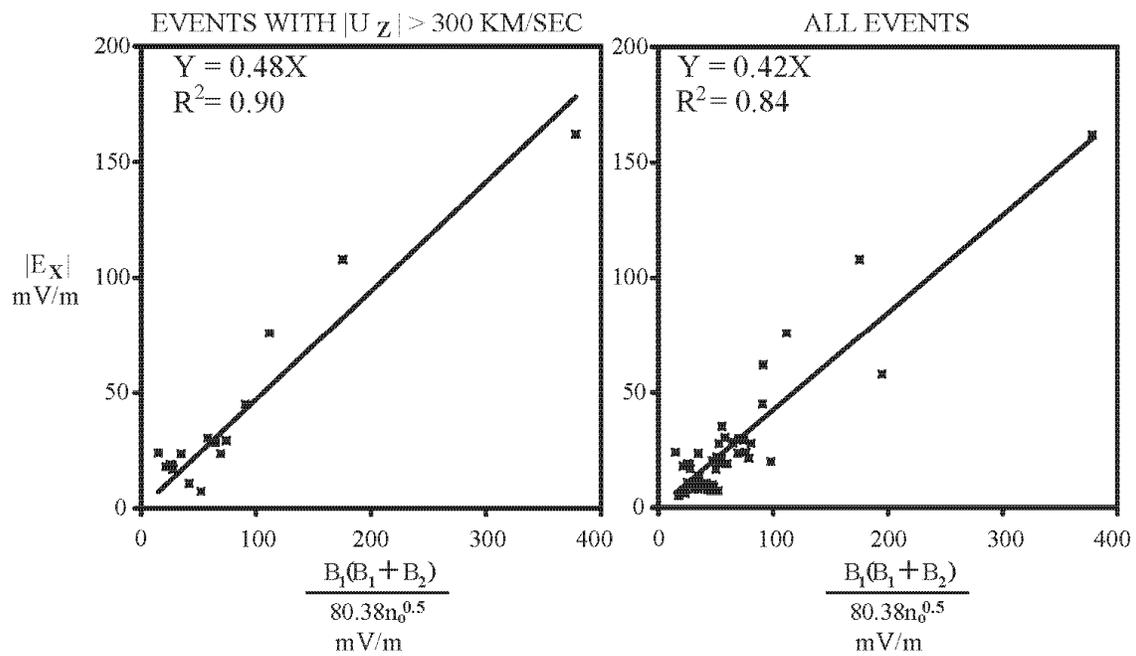

Figure 3